\documentclass{appolb}
\usepackage{epsfig}
\usepackage{float}

\begin{document}
\pagestyle{plain}
\newcount\eLiNe\eLiNe=\inputlineno\advance\eLiNe by -1
\title{Competing contact processes on homogeneous networks with tunable clusterization
\thanks{Correspondence to: {\tt kulakowski@fis.agh.edu.pl}}
}
\author{ Marcin Rybak and Krzysztof Ku{\l}akowski
\address{Faculty of Physics and Applied Computer Science, AGH University of Science and Technology, al. Mickiewicza 30, PL-30059 Krak\'ow, Poland
}}

\maketitle
\begin{abstract}
We investigate two homogeneous networks: the Watts-Strogatz network and the random Erdos-Renyi network, the latter with tunable clustering coefficient $C$. The network is an area of two competing contact processes, where nodes can be in two states, S or D. A node S becomes D with probability 1  if at least two its mutually linked neighbours are D. A node D becomes S with a given probability $p$ if at least one of its neighbours is S. The competition between the processes is described by a phase diagram, where the critical probability $p_c$ depends on the clustering coefficient $C$. For $p>p_c$ the rate of state S increases in time, seemingly to dominate in the whole system. Below $p_c$, the contribution of D-nodes remains finite. The numerical results, supported by mean field approach, indicate that the transition is discontinuous.
\end{abstract}
  
\section{Introduction}

Structure and dynamics of communication networks is a key issue in areas as different as modern ecology \cite{ec1,ec2}, sociology \cite{sn1,sn2} and computer sciences \cite{cn}. Statistical physics started its massif and interdisciplinary venture with networks only recently \cite{ws,ab}, bringing there an interdisciplinary attitude, the search of phase transitions and of universal laws. However, the terrain was well prepared by tens of years of research on lattices. In this approach percolation, cellular automata and contact processes are keywords, and the key method is computer simulation \cite{prc,ca,cp}. \\

Here we are interested in contact processes on networks with tunable clustering. This interest is by no means new; actually, the problem is actively 
investigated in terms of percolation \cite{gl1,gl2,gl3,gl4} and competing diseases \cite{di1,di2,di3,di4,di5}. In a recent paper \cite{di4}, the case of two diseases has been considered, with mutual interaction of diseases through cross-immunity. The difference between the diseases was only in the values of their transmissibilities. Our aim here is to investigate two competing processes of different nature: one is activated by one neighbor, and the other - by a connected pair of two neighbors. In terms of opinion spreading, the problem could be termed as Voter vs Sznajd model \cite{legtt,sznaj}. Up to our knowledge, this competition has not been investigated yet. As expected, the tunable clustering is particularly relevant for the process activated by connected pairs. According to this view, the Sznajd dynamics (activation by pairs) itself should strongly depend on the clustering coefficient. This comment was given already in \cite{sou}, but the structure dependence on the dynamics was not considered there. Here we assume that a node activated by a pair is necessarily the nearest neighbor of each node of the activating pair. This is different than in the Sznajd model, where all neighbors of at least one node of this pair are activated. Further, a pair activates one site in one step. Making these modifications, we are motivated by sociological theory of the threshold effect \cite{gran,mass,chwe}. Although this theory is sometimes brought up in the literature on the Sznajd model \cite{ksw}, in that approach a node is activated being a neighbor of only one member of the pair, what is not consistent with the sociological mechanism of the threshold effect. On the other hand, dynamics of collective social effect was modelled with success within the Random Field Ising model \cite{bou}. Still, up to our knowledge, a competition of dynamics activated by two different processes has not been investigated yet. \\

In our considerations, we distinguish between the change activated by one and at least two neighbours. Two is a minimal version of the threshold mechanism different from the Voter model; often, the number of neighbours necessary to activate a given behaviour is proposed to be larger than two. The difference between dyads and triads in social contagion is accepted in literature \cite{mm} and it was underlined already by Georg Simmel \cite{sim}. Our problem can be appllied in modeling these kinds of behaviour which need two persons to be imitated; in fact, threshold of two or more persons was found in social measurements \cite{tva}. In particular, statistics on marriages and divorces deserves attention, because the imitation mechanism can be active there and at it can be least partially responsible for the observed effects. In this case, the competing processes are the imitation of married vs divorced persons. The list of examples of behavior activated by pairs of persons include also joining scuffles and demonstrations of sexuality. To mention other possible applications, the model can serve as a step towards simulation of inheriting some genetic ilnesses, as hemophilia, where the presence of defected genes in one or two parents does matter \cite{wei}. Further, an application of the model is possible in computer sciences, in the so-called Paxos family of consensus protocols \cite{pax}. Namely, suppose that a part of code or data in neigboring units is particularly exposed to damages, and the positions of these damages are random. Then, the data can be repaired by a comparison of the datasets in two neighboring units, because it is unlikely that data in two units are damaged at the same position. Finally, our model can be relevant in economy, where different behaviour is observed of dyads and triads of firms \cite{eko1,eko2}, with probability of collusions as an example.\\

The text is organized as follows. The model is explained and our numerical results are shown in Section II and III, respectively. In Section IV the related mean field approach is presented.  The comparison of these results is discussed in the last section.

\section{The model}

The medium for our simulation, homogeneous networks, are either the Watts-Strogatz network or the Erd\"os-R'enyi network. In the former, the clustering coefficient $C$ is known to depend on the rewiring probability $\beta$ \cite{ws}. The latter networks are prepared from the usual Erd\"os-R'enyi network by adding some links between neighbors of the same nodes. The procedure, applied by Holme and Kim to the scale-free networks \cite{hol}, was adapted later \cite{amk} to the Erd\"os-R'enyi networks. As a rule, the modified networks contained $10^3$ nodes, with mean degree $<k>=10.00\pm0.5$.  Note that the degree distribution for the networks is different than the Poison distribution \cite{amk}. The clustering coefficient $C$ is calculated for each network.\\

All nodes are labeled with S or D. Initially, some amount of nodes are selected to be D-nodes, with all remaining nodes marked as S-nodes. The dynamic rules are as follows: each S-node, which has at least two mutually connected D-nodes as nearest neighbors, becomes D-node with probability 1, and each D-node which has at least one S-node as nearest neighbor, becomes S-node with probability $p$. Actually, what is relevant is the ratio of these rates, $p_S/p_D=p/1$, and the probability of activation of D-nodes is fixed as one for convenience. The nodes are modified in parallel: nodes S at even steps, nodes D at odd steps. The advantage of this way of updating is that the results are more stable in time. The drawback is that we get two curves, one after updating S-nodes and another after D-nodes, and sometimes the curves do not merge. The difference between the curves can be treated as a measure of numerical accuracy. \\

Outcome of the simulation is to find the contributions of S-nodes as dependent on the probability $p$ and the clustering coefficient $C$. The latter 
itself depends on the rewiring probability $\beta$ for the WS networks and on the number of added links for the ER networks. We find that the observed percentage of S-nodes (and of course of D-nodes) shows some discontinuities. This allows to present the results in the form of phase diagrams.
 
\section{Numerical results}
 
In Fig. 1 we show time dependences of the percentage $x_S=N_S/N$ of S nodes for an exemplary values of the clustering coefficient $C$ for the Watts-Strogatz network. The curves can be divided into two parts, qualitatively different. For probability $p$ smaller than some value $p_c(C)$, $x_C$ becomes constant after some transient time. For $p>p_c$, $x_S$ slowly increases, seemingly to the maximal value 1.0. As a rule, the curves representing two kinds of updating merge. In Fig. 2, the curve $p_c(C)$ is shown. The results indicate that $x_S$ is discontinuous at $p_c$; this is shown in Fig. 3. We note that for $p=0$, the results on $x_S$ strongly depend on the initial state, whilst they do not for $p>0$. \\

The same kind of results for the Erd\"os-R'enyi networks are shown in Figs. 4-6. In this case, the two updating curves merge only below $p_c$. Above this threshold, the split between the curves remains finite. 
 
\begin{figure}[H] 
\vspace{0.3cm} 
 \includegraphics[angle=270,width=9cm]{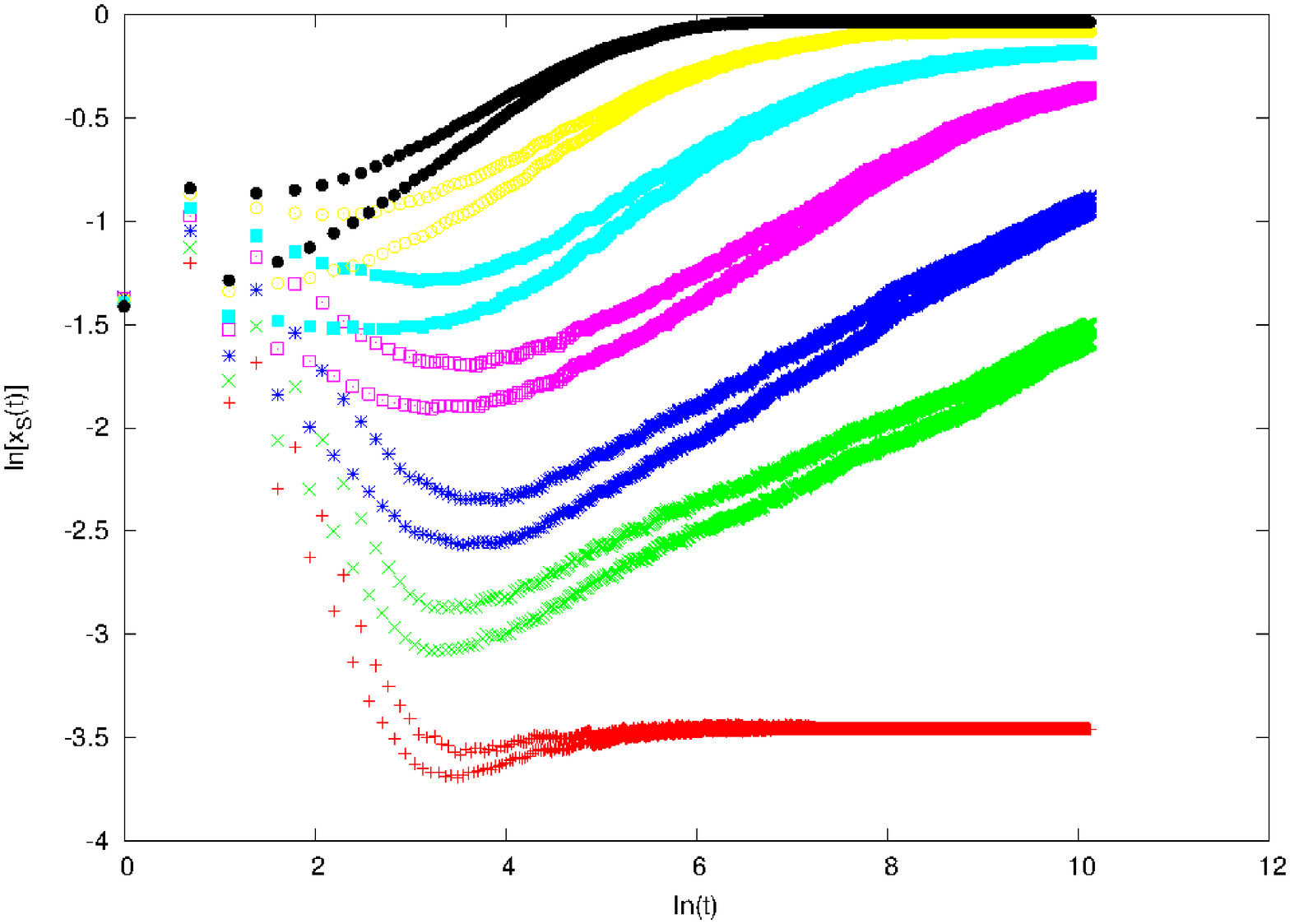} 
 \includegraphics[angle=270,width=9cm]{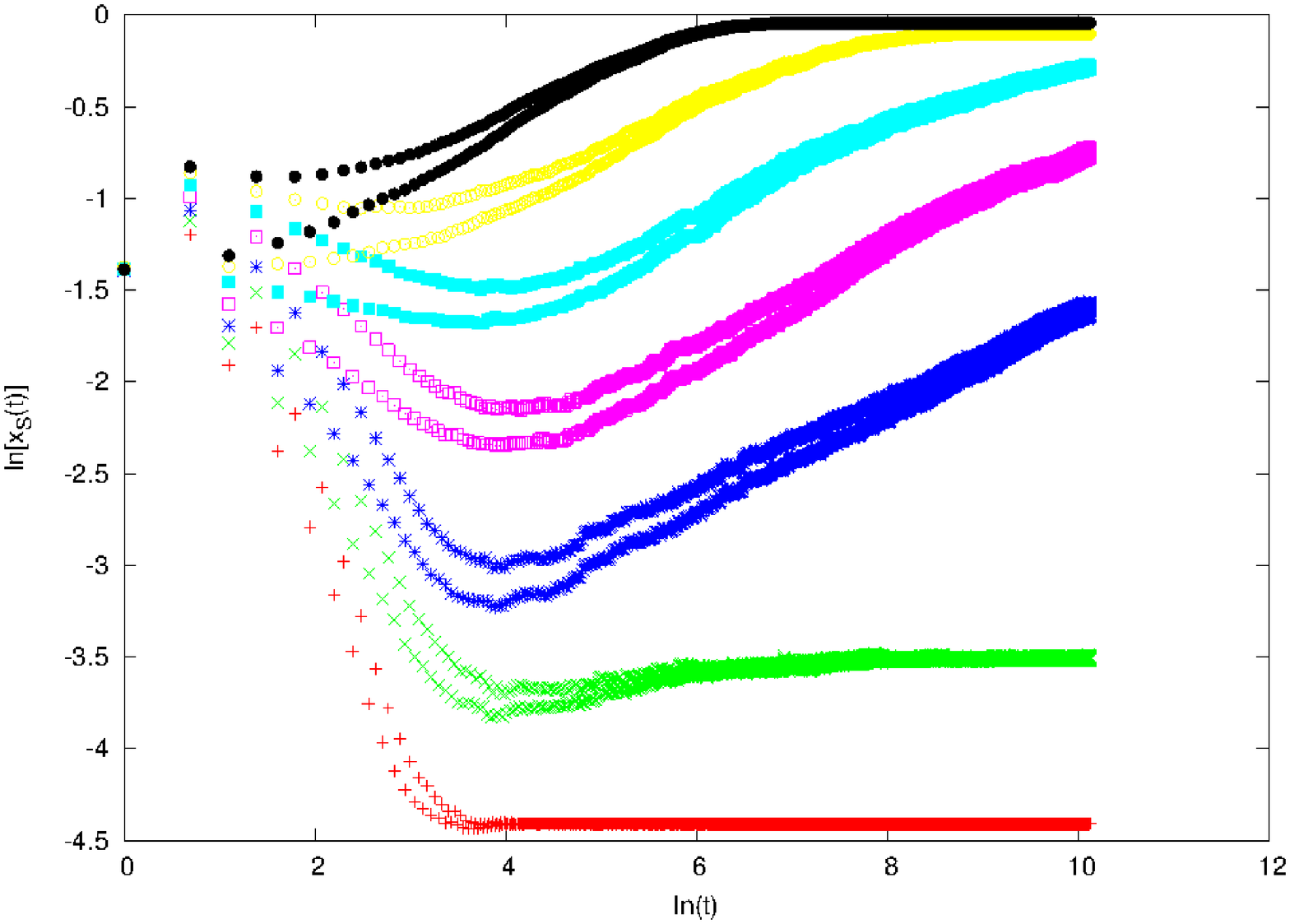} 
 \includegraphics[angle=270,width=9cm]{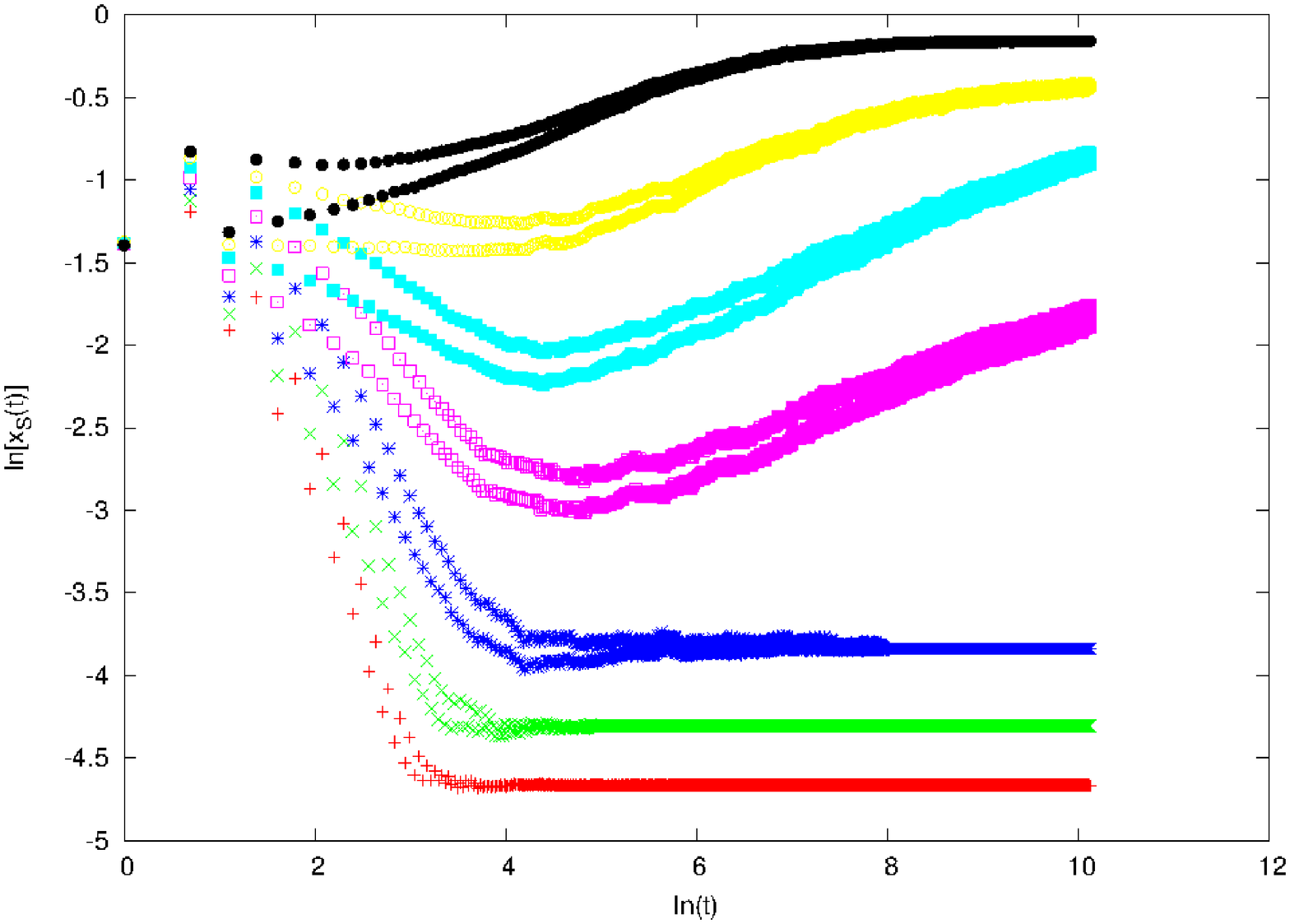} 
 
\vspace{0.3cm} 
\caption{Time dependence of the S-nodes participation in all nodes amount for the Watts-Strogatz network (25\% of all 1000 nodes are S-nodes initially). Each graph represents different clustering coefficient value: 0.44 for the top graph, 0.45 for the middle and 0.47 for the bottom one.
Every curve on the each graph, has assigned probability value from range 0.1 to 0.4 with step 0.05, in ascending order from the bottom to the top of the graph.}  
\label{fig-1}
\end{figure} 

\begin{figure} 
\vspace{0.3cm} 
 \includegraphics[angle=270,width=10cm]{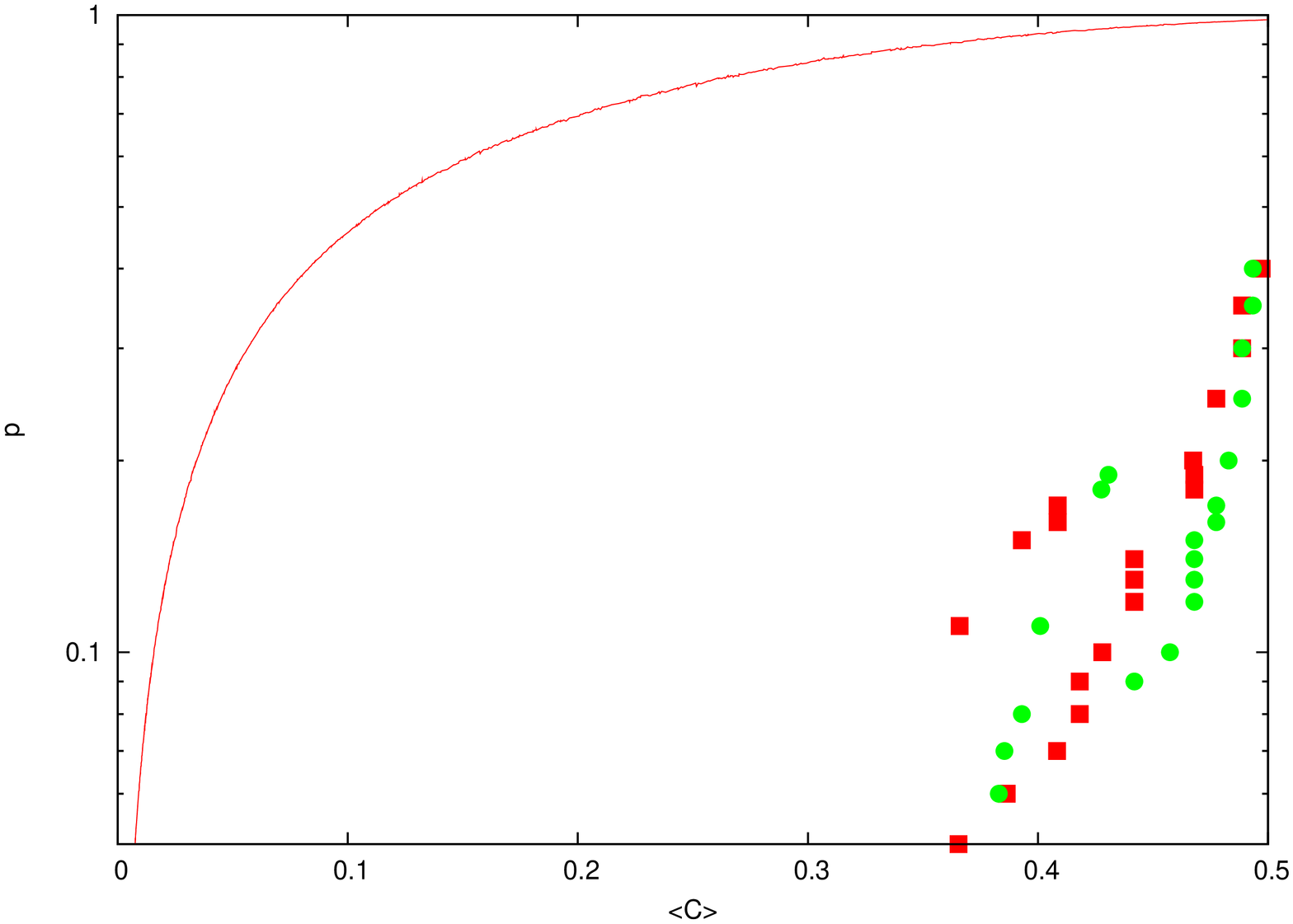} 
 
\vspace{0.3cm} 
\caption{Dependence between critical probability and clustering coefficient for different initial S-nodes participations (red squares: 25\% of all 1000 nodes, green circles: 75\% of all 1000 nodes), in relation to mean field theory for the Watts-Strogatz network. The solid line represents f(C) function which is discussed in Chapter 4.  } 
\label{fig-2}
\end{figure} 

\begin{figure}[H] 
\vspace{0.3cm} 
 \includegraphics[angle=270,width=10cm]{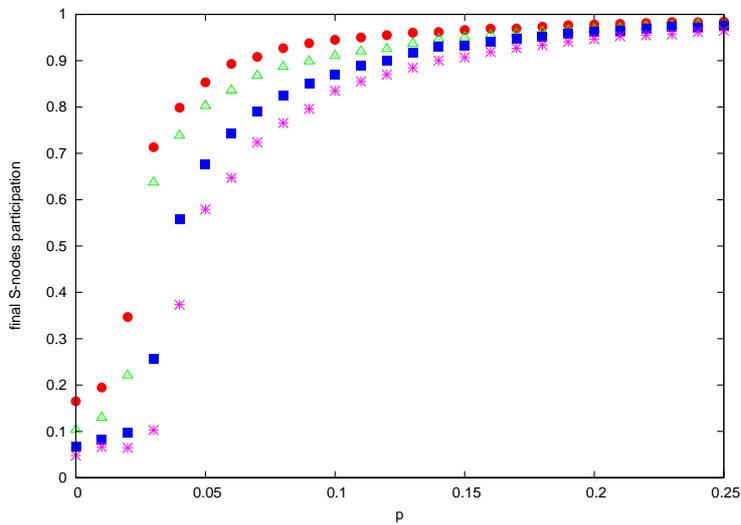} 

\vspace{0.3cm} 
\caption{Dependence between final S-nodes participation  and probability for the Watts-Strogatz network (25\% of all 1000 nodes are S-nodes initially) for different clustering coefficient values: circles: 0.11, triangles: 0.21, squares: 0.27, stars: 0.31. }  
\label{fig-3}
\end{figure}

\begin{figure}[H] 
\vspace{0.3cm} 
 \includegraphics[angle=270,width=9cm]{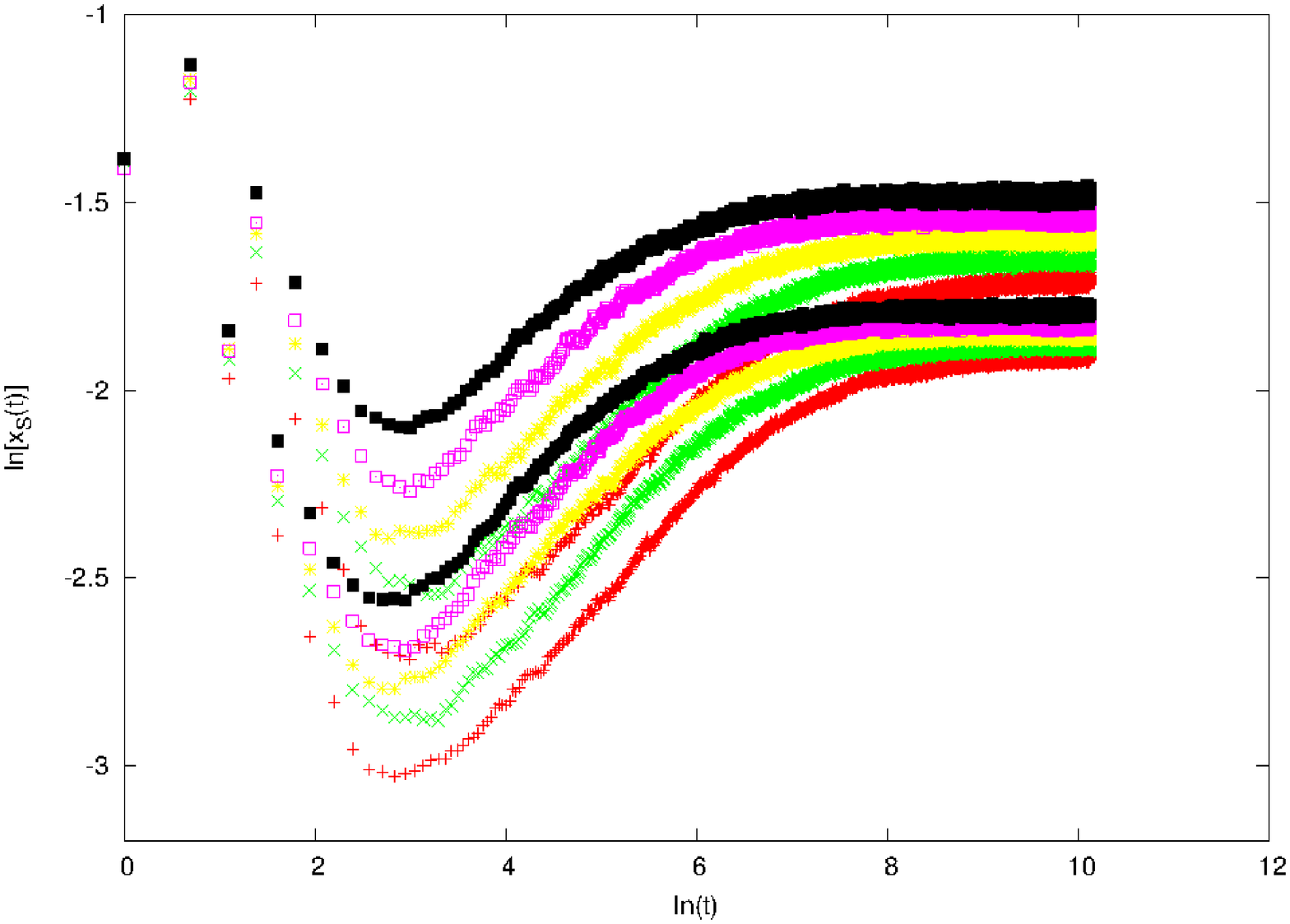} 
 \includegraphics[angle=270,width=9cm]{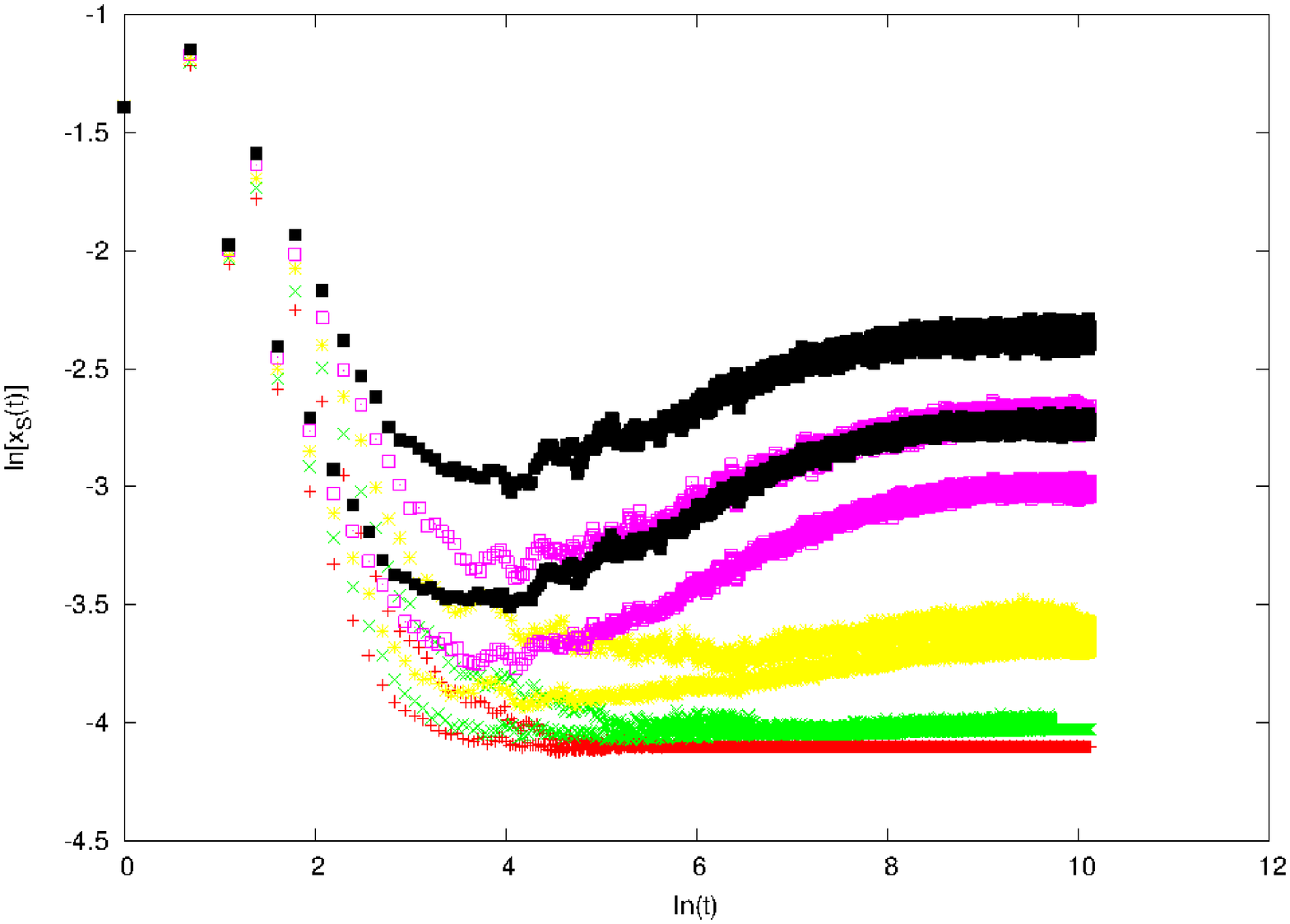} 
 \includegraphics[angle=270,width=9cm]{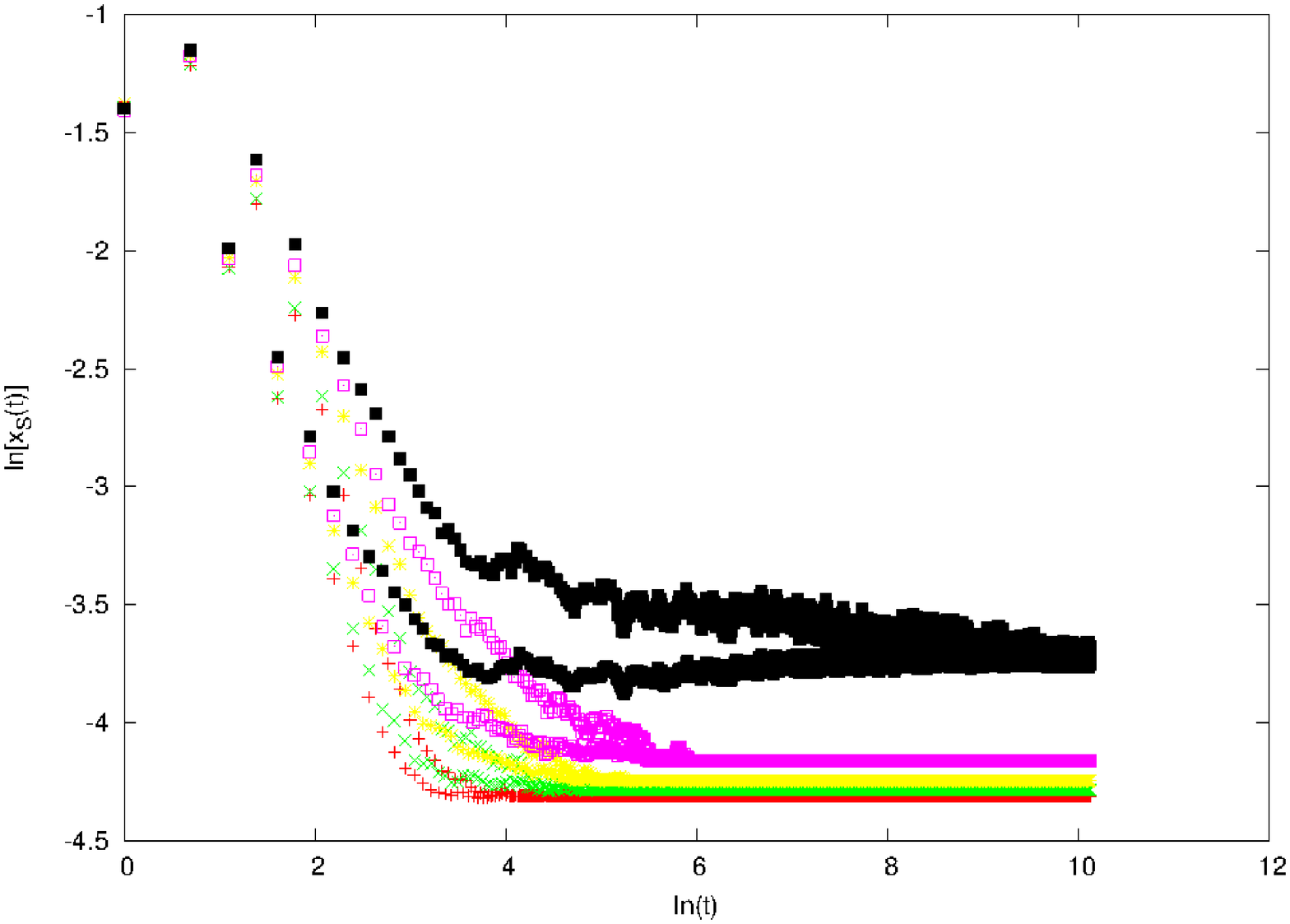} 
 
\vspace{0.3cm} 
\caption{Time dependence of the S-nodes participation in all nodes amount for the Erdos-Renyi network (25\% of all 1000 nodes are S-nodes initially). Each graph represents different clustering coefficient value: 0.1 for the top graph, 0.19 for the middle and 0.29 for the bottom one.
Every curve on the each graph, has assigned probability value from range 0.07 to 0.11 with step 0.01, in ascending order from the bottom to the top of the graph.}  
\label{fig-4}
\end{figure} 

\begin{figure}[H] 
\vspace{0.3cm} 
 \includegraphics[angle=270,width=10cm]{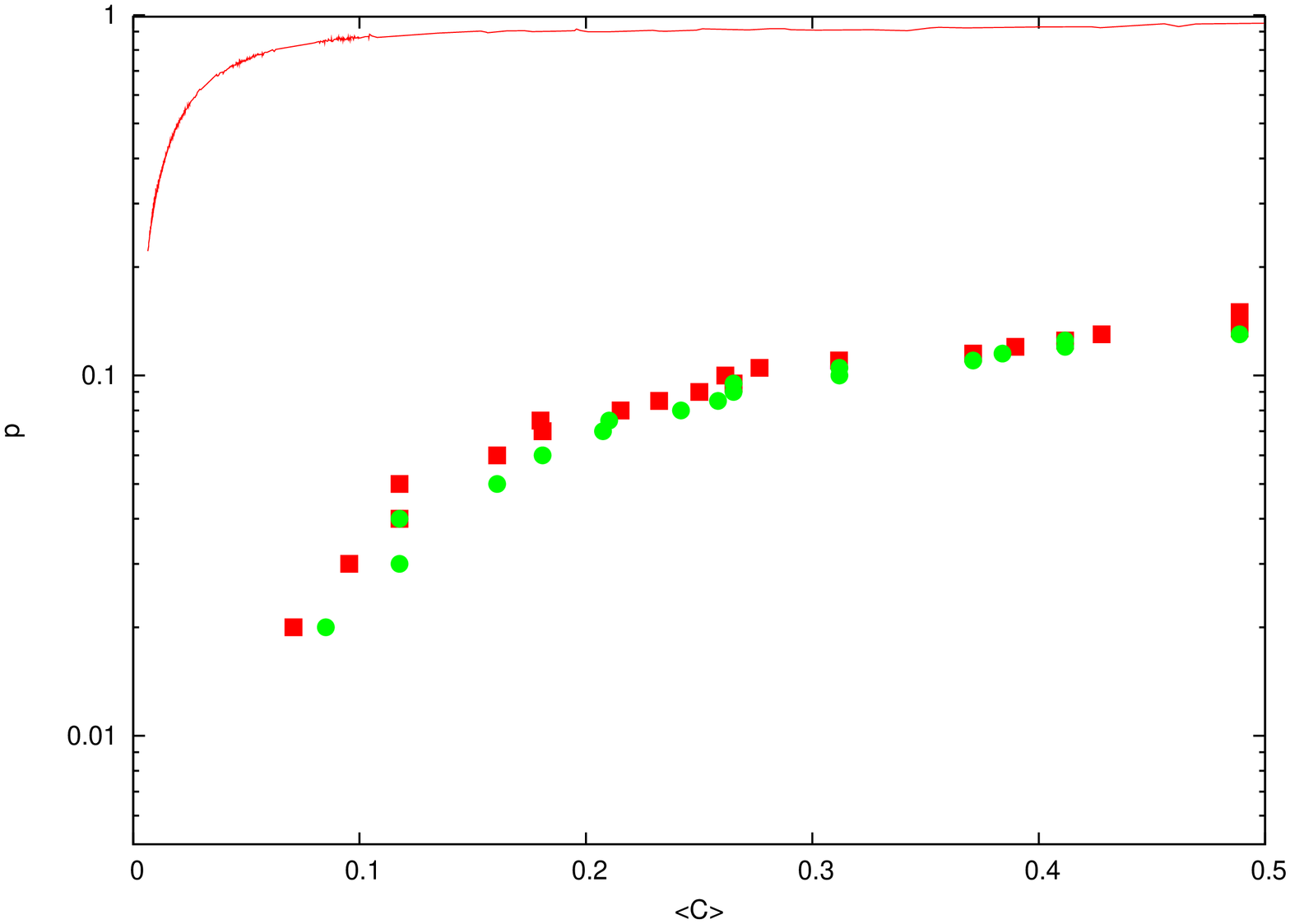} 
 
\vspace{0.3cm} 
\caption{Dependence between critical probability and clustering coefficient for different initial S-nodes participations (red squares: 25\% of all 1000 nodes, green circles: 75\% of all 1000 nodes), in relation to mean field theory for the Erdos-Renyi network. The solid line represents f(C) function which is discussed in Chapter 4.  } 
\label{fig-5}
\end{figure} 

\begin{figure}[H] 
\vspace{0.3cm} 
 \includegraphics[angle=270,width=10cm]{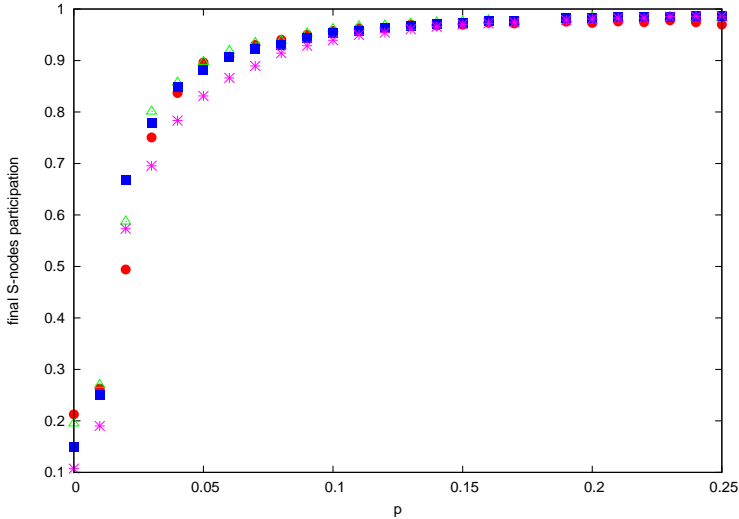} 

\vspace{0.3cm} 
\caption{Dependence between final S-nodes participation  and probability for the Erdos-Renyi network (25\% of all 1000 nodes are S-nodes initially) for different clustering coefficient values: circles: 0.006, triangles: 0.01, squares: 0.02, stars: 0.03. } 
\label{fig-6}
\end{figure}

\section{Mean field}

To interpret the numerical results, it is convenient to compare them with related mean-field theory.  In the formalism of Master equations \cite{grd}, the time derivative of the percentage $y$ of D-nodes can be written as

\begin{equation}
\dot y = f(C)xy^2-pxy
\end{equation}
where $x=1-y$ is the percentage of S-nodes, and $f(C)$ is the probability that a node has at least two, mutually connected neighbors. It is easy to see that for small values of the clustering coefficient $C$ the function $f(C)\propto C$. The argument is as follows: for the degree distribution $P(k)$ the probability that a node has at least two neighbors is $P(2)+P(3)+...=1-P(0)-P(1)$. The probability $f(C)$ that at least one pair of these neighbors is connected is $P(2)C+P(3)[1-(1-C)^3]+P(4)[1-(1-C)^6]+...$; in general, it is equal to

\begin{equation}
f(C)=\sum_{k=2}^{\infty}P(k)[1-(1-C)^{k(k-1)/2}].
\end{equation}
The factor $[...]$ at $P(k)$ means that it is not true that any pair of neighbors is not mutually linked. For small $C$, this factor can be approximated as $k(k-1)C/2$, and we get $f(C)=(<k^2>-<k>)C/2$,  positive if $C$ is not zero. \\

It is clear that Eq. 1 is controlled by one parameter $a=p/f(C)$. There are three fixed points: $y*=0$ is always stable, $y*=1$ is stable in the range $0<a<1$ and unstable for $a>1$, where $y*=a$ becomes stable. However, this transcritical bifurcation is not of interest for us, because $y>1$ is meaningless. Similarly, the fixed point $y*=a$ is unstable in the range $0<a<1$. The case when $p=0$ is to be considered separately; then, the only stable solution is $y*=2/3.$\\

Summarizing, for $p>0$ we have to distinguish two cases. If $p>f(C)$, the only stable solution is $y*=0$, i.e. all nodes are S-nodes. If $p<f(C)$, the final state depends on the initial one and we have to distinguish two sub-cases. If initially $y<p/f(C)$, the final state is again all S-nodes. If, on the contrary, $y>p/f(C)$ in the initial state, all nodes become D-nodes.\\ 

The solution is similar, if we withdraw the condition that activating a S-node must be accompanied by a S-neighbor. In this case the last term in Eq. 1 is just $-py$. Here again, the fixed point $y*=0$ is always stable. Two other fixed points exist below the saddle-node bifurcation at $p/f(C)=1/4$; the higher one ($y*_{+}=(1+\sqrt{1-4a})/2$) is stable, the lower one ($y*_{-}=(1-\sqrt{1-4a})/2$) is unstable. This means again, that for $p>f(C)/4$ all nodes end as S-nodes, but the choice $p<f(C)/4$ and the initial concentration of D-nodes larger than $y_{-}$ should lead to all D-nodes in the final state.\\

\section{Discussion}

The comparison of numerical results with their analytical counterparts allows to identify the similarities and differences. The analytical result for $p=0$ is that for any value of the clustering $C$, the S nodes appear in 2/3 of the system. According to the numerical result, for $p=0$ there is a strong dependence of $x_C$ on the clustering coefficient $C$ and even on an initial state. Further, both methods indicate that for $p>0$, there is a critical value of $p$ (denoted here as $p_c$) such that $x_S$ is one (analyticals) or tends to one (numericals) for $p>p_c$. However, the results are different for $p<p_c$. The analytical solution is that $x_S$ is 0 or 1, depending on the initial state ($x_S$ smaller or larger than $f(C)/4$, respectively, where $f(C)$ is given by Eq. 2). The numerical solution is that $x_S$ depends on $C$. Some weak dependence on the initial state can be due to numerical inaccuracy. Despite this kind of errors, the numerical method captures an influence of local correlations and therefore is more credible.\\

When related to the imitation and the threshold effect in sociology, our results indicate that the static outcome changes discontinuously as dependent on the probability $p$ of activation of S-nodes. In this application, D-nodes denote persons who join an activity according to the threshold, and S-nodes denote persons who do not. Keeping the example of divorcing, our algorithm is to be interpreted as imitation of married couples and divorced persons. The
discontinuity, which is our main results of qualitative significance, means that in a stationary phase the number of divorced persons changes abruptly. More directly, an increase of number of divorces is related to an increase of the rate of activation of S-nodes (singles) with respect to the activation of D-nodes (married). This kind of effect should be readable in statistical data. Up to our knowledge, most sharp relative change of the number of divorces happened in Portugal during the so-called Carnation Revolution, when the statistics show a jump from 700 to 7000 divorces per year between 1973 and 1977 \cite{port}. However, in that case the possible effect of imitation is hidden under the consequences of liberalization of law. A better example is the increase of number of divorces in United States by about 40 percent, starting about 1965 \cite{usea}. The process can be associated with devastating consequences of Vietnam war, and certainly it was initialized by it. However, a similar increase just after WWII quickly disappeared, in contrast to the persistence of the effect in 70's. On the other hand, an increase of divorces in England and Wales around 1970 \cite{enwal} cannot be interpreted as a consequence of an armed conflict, but rather as a demonstration of a more general phenomenon which had happened with various speed in most European countries \cite{usea}. There, we feel entitled to ask for a role of imitation. \\

Concluding, our results show a sharp dependence of the final state on the probability of activation of simulated process. This sharp dependence 
can be related to competing processes of imitation in sociology. Other applications in behavioral economy, immunology and computer sciences are also possible.\\
 
\section*{Acknowledgements} The authors are grateful to Michael Macy, Andrzej Maksymowicz, Luzius Meisser and Matthias Meyer for helpful comments and providing reference data. The calculations were performed in the ACK Cyfronet, Cracow, grants No. MNiSW/IBM\_BC\_HS21/AGH/070/2010 and MNiSW/SGI3700/ AGH/070/2010. 
This work was partially supported from the AGH UST project No. 10.10.220.01.

\end{document}